# A Novel APVD Steganography Technique Incorporating Pseudorandom Pixel Selection for Robust Image Security

Mehrab Hosain[1][0009-0007-1079-1052] and Rajiv Kapoor[2][0000-0003-3020-1455]

[1] Delhi Technological University, Delhi 110042, India
robinhosain@gmail.com
[2] Delhi Technological University, Delhi 110042, India
rajivkapoor@dce.ac.in

**Abstract.** Steganography is the process of embedding secret information discreetly within a carrier, ensuring secure exchange of confidential data. The Adaptive Pixel Value Differencing (APVD) steganography method, while effective, encounters certain challenges like the "unused blocks" issue. This problem can cause a decrease in security, compromise the embedding capacity, and lead to lower visual quality. This research presents a novel steganographic strategy that integrates APVD with pseudorandom pixel selection to effectively mitigate these issues. The results indicate that the new method outperforms existing techniques in aspects of security, data hiding capacity, and the preservation of image quality. Empirical results reveal that the combination of APVD with pseudorandom pixel selection significantly enhances key image quality metrics such as Peak Signal-to-Noise Ratio (PSNR), Universal Image Quality Index (UIQ), and Structural Similarity Index (SSIM), surpassing other contemporary methods in performance. The newly proposed method is versatile, able to handle a variety of cover and secret images in both color and grayscale, thereby ensuring secure data transmission without compromising the aesthetic quality of the image.

## 1   Introduction

The rapid growth of the internet and the massive exchange of sensitive data in the digital era have increased the importance of secure communication. Steganography plays a crucial role in this regard, as it offers a way to hide secret information within a carrier medium, such as images, audio, or video files, ensuring that the concealed data is undetectable [2, 6]. Unlike cryptography, which focuses on securing the message through encryption, steganography aims to make the message invisible to eavesdroppers by embedding it within a seemingly innocent medium [19, 30] Image steganography has gained substantial attention due to its widespread application and the vast availability of digital images [6, 11, 12]. Various image steganography techniques have been proposed, including the Least Significant Bit (LSB) substitution method, which is widely employed due to its simplicity and ease of implementation [3, 15, 21]. However, LSB-based methods are vulnerable to steganalysis attacks, as they often introduce detectable artifacts in the stego-image [7, 27]. Pixel Value Differencing (PVD) is another popular image steganography technique that exploits the difference between adjacent pixel values to embed secret data [16, 31]. Adaptive Pixel Value Differencing (APVD) techniques have been developed to address the drawbacks of traditional PVD methods, such as the "fall off boundary problem" [16, 18, 31]. However, APVD methods still encounter the "unused blocks" problem, which can impact the efficiency and visual quality of the stego-image [23].

In this paper, Introduce a novel image steganography approach that combines the advantages of APVD with pseudorandom pixel selection to effectively address the drawbacks of APVD methods [16, 18, 31]. The technique propose offers significant improvements in security, expands the data embedding capacity, and provides a notable improvement in visual quality compared to other techniques. These techniques include methods such as LSB matching, deep learning-based steganography, Discrete Cosine Transform (DCT), APVD, traditional PVD, quantum substitution boxes, and traditional LSB methods [3, 7, 13, 22].

This research is organized as follows: Starting with a review of the existing literature in the field, it progresses to explaining the newly proposed APVD with Pseudorandom pixel selection methodology. Then, the data used in this investigation is presented, followed by the experimental results and their respective evaluations. Conclusions based on these findings are drawn, and finally, possible future research directions are outlined

## 2  Related Work

Recent developments in image steganography techniques have aimed at enhancing the security, robustness, and capacity of these methods. As shown in Table 1, several studies have explored different techniques to achieve these objectives. In the area of image steganography, a several of techniques and approaches have been recommended in recent years. Luo et al. [16] proposed an Adaptive Pixel Value Differencing (APVD) scheme to enhance security and visual quality compared to traditional PVD techniques. APVD adjusts the embedding capacity based on the cover image characteristics, striking a balance between imperceptibility and capacity. However, APVD still faces the "unused blocks" problem, that could potentially influence the general efficacy and aesthetic appeal of the stego-image. Ahmed A. Abd et al. [1] proposed a novel image steganography technique was proposed using quantum walks for secure S-boxes. While the method demonstrated improvements in security and visual quality, it has certain limitations, such as the reliance on complex quantum algorithms, which may hinder its practical application in some scenarios. Tang et al. [25] presented an automatic steganographic distortion learning method using a generative adversarial network (GAN), which improved the security of steganographic systems. He et al. [9] introduced a reversible data hiding technique using multi-pass pixel value ordering and prediction-error expansion, which improved the payload capacity and image quality. Yang et al. [28] proposed a deep learning-based method to hide an image within another image, improving the robustness of the hidden data against attacks. Yang et al. [29]

presented an efficient color image encryption method by color-grayscale conversion based on steganography, offering increased security for color images. Deng et al. [4] developed a universal image steganalysis approach based on convolutional neural networks with global covariance pooling, which improved the detection performance of hidden messages in various steganographic methods. Tseng and Leng [26] proposed a reversible modified least significant bit (LSB) matching revisited method that increased the robustness and reversibility of hidden data. Tang et al. [24] broke new ground by developing an adaptive fuzzy inference technique specifically for color image steganography. Their novel approach significantly bolstered the security of hidden data, while also improving its imperceptibility. Khare [13] brought a fresh perspective to the field by introducing a unique image functioned framework that applied the PS-DCT (Phase Shift - Discrete Cosine Transform) technique for steganography. This groundbreaking approach notably enhanced the imperceptibility of the hidden data while simultaneously increasing the payload capacity. Pandey et al. [17] unveiled a highly effective strategy for secret data transmission that combined advanced steganography with image compression. This innovative method offered marked improvements in the efficiency and security of data transmission, further enriching the field of steganography. Finally, Frățilă and Morogan [5] explored enhanced models in deep image steganography, aiming to improve the security and capacity of steganographic systems using deep neural networks. As shown in Table 1, several studies that have delved into different techniques, yet some limitations have been identified.

**Table 1.** Relevant work in images steganography

| Method | Weakness |
| --- | --- |
| APVD [16] | Unused blocks problem |
| Quantum S-Box [1] | Complex algorithms for practical application |
| GAN [25] | High complexity |
| Reversible data hiding [9] | Limited to specific image formats |
| Deep learning [28] | Requires large training datasets |
| Color-grayscale conversion Based [29] | May introduce color distortion |
| CNN with global covariance pooling [4] | Limited to specific steganographic techniques |
| Reversible LSB matching [26] | Possible security gaps beyond RS-analysis. |
| Fuzzy inference [24] | Complexity in implementation |
| PS-DCT approach [13] | Limited robustness against more complex steganalysis techniques. |
| Advanced steganography and image compression [17] | Complexity in implementation |
| Deep neural networks [5] | Requires large training datasets |

# 3 Proposed Methodology

The Proposed Methodology section lays out the innovative approach developed in this study, which integrates Adaptive Pixel Value Differencing (APVD) with a pseudorandom pixel selection strategy. The overall objective of this approach is to fortify image security while maintaining quality, and to effectively address the "unused blocks" challenge observed in existing APVD methods.

## 3.1 Adaptive Pixel Value Differencing (APVD)

This method offers an effective and versatile approach to image steganography, delivering enhanced security, higher embedding capacity, and better visual quality preservation [16, 18]. By adaptively modifying pixel value differences and leveraging the natural redundancy in images, APVD makes hidden data more difficult to detect while accommodating larger secret messages.
The APVD method determines the data embedding capacity for each pixel pair by taking into account the differences in their pixel values. For a pixel pair ($P_1$, $P_2$), the pixel value difference, D, is computed as follows:

$$D = |P2 - P1| \qquad (1)$$

The data embedding capacity for this pixel pair is ascertained based on the range in which the difference (D) is located, Eq. (1). The relationship between the range of D and the embedding capacity can be expressed as follows eq. (2):

$$k = f(D) \qquad (2)$$

In Eq. (2), $k$ is the number of bits that can be embedded and the function $f$ maps the value of D to the associated embedding capacity. By adapting the embedding capacity based on the pixel value difference, APVD ensures efficient utilization of the available pixel pairs, accommodating larger secret messages while maintaining visual quality.

## 3.2 Pseudorandom Pixel Selection

Pseudorandom pixel selection shows a crucial role in improving the security and visual imperceptibility of the steganography process. It involves utilizing a pseudorandom number generator (PRNG) to determine the order in which the pixel pairs are selected for embedding and extraction [25], [26]. To ensure successful extraction of the secret data, the PRNG is initialized with a shared seed value, denoted PRNG(Sseed):

$$PRNG(S_{seed}) \rightarrow P_i \qquad (3)$$

Eq. (3), Pi represents the ith pixel pair in the pseudorandom sequence. To generate the pseudorandom sequence of pixel pairs, the PRNG function is applied iteratively, as follows:
$P_1 = PRNG(S_{seed})$

$$P_2 = PRNG(P_1)$$
$$P_3 = PRNG(P_2) \quad (4)$$
$$P_n = PRNG(P_{(n-1)})$$

where 'n' represents the cumulative count of pixel pairs present in the image. By using this pseudorandom sequence Eq. (4), the pixel pairs are selected for embedding and extraction in a more secure and visually imperceptible manner.

### 3.3 APVD with Pseudorandom Pixel Selection

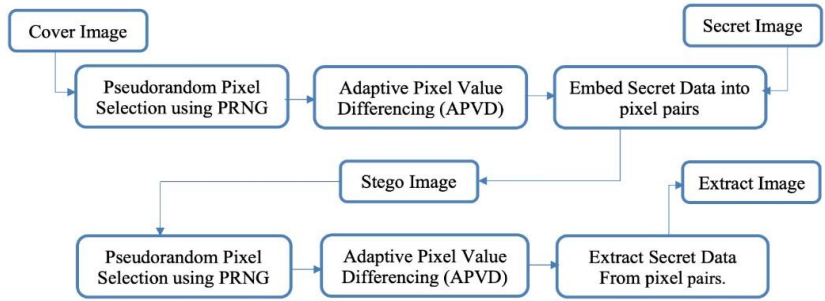

**Fig. 1**. Diagrammatic Representation of Integration and Retrieval Procedures

Visual representation of both embedding and extraction procedures integral to the suggested image steganography method can be perceived in the supplied Fig. 1. The onset of the embedding process involves treating the cover image with a pseudorandom pixel selection operation, enabled through a pre-agreed seed value. Subsequently, the chosen pixel pairs undergo the technique of Adaptive Pixel Value Differencing (APVD) where the covert information is seamlessly integrated within these pairs. Post this process, the modified pixel pairs congregate, leading to the formation of the stego-image. In a contrasting manner, the extraction phase initiates by processing the stego-image with the pseudorandom pixel selection operation once more, utilizing the identical shared seed value to maintain the pixel pair selection order. These pixel pairs are then subjected to the APVD operation again, although in this instance, covert data is retrieved from within the pixel pairs based on their value differences. This process concludes when the retrieved data fragments are assembled, thereby reproducing the initial covert message and marking the successful completion of the extraction procedure.

**Embedding Process:**
1. Initialize the PRNG with the shared seed value $S_{seed}$.
2. For i = 1 to n, where 'n' represents total count of pixel pairs in the stego-image:
   a) Use the PRNG to select a pixel pair ($P_1$, $P_2$) pseudo-randomly: $(P1, P2) = PRNG(Sseed, i)$
   b) Calculated the discrepancy in pixel value, denoted as D, for the chosen pairs of pixels: $D = |P2 - P1|$

  c) Establish the pixel pair's data-embedding potential, symbolized by k, using the range of D as a basis.
  d) Incorporate k bits of concealed information into the pixel pair through appropriate alterations in pixel values: $(P1', P2') = Embed\ (P1, P2, s)$.
3. Combine the modified pixel pairs ($P_1'$, $P_2'$) to form the stego-image.

**Extraction Process:**
1. Initialize the PRNG with the shared seed value $S_{seed}$.
2. For i = 1 to n, where n represents the total count of pixel pairs in the stego-image:
   a) Use the PRNG to select a pixel pair ($P_1'$, $P_2'$) pseudo-randomly from the stego-image: $(P1', P2') = PRNG(Sseed, i)$
   b) Compute the pixel value difference D for the selected pixel pairs: $D = |P2' - P1'|$
   c) Determine the embedding capacity k of the pixel pair based on the range of D.
   d) Extract k bits of secret data s from the pixel pair by reversing the modifications made during the embedding process: s = Extract ($P_1'$, $P_2'$, D)

Merge the unveiled secret data elements to recreate the original concealed message.

### 3.4  Color Image Handling

For color images, the APVD method can be applied independently to each of the RGB channels. Let $P_1 = (R_1, G_1, B_1)$ and $P_2 = (R_2, G_2, B_2)$ be the pixel pairs in the RGB channels [2]. The pixel value differences for each channel are calculated as follows:

$D_R = |R_2 - R_1|$

$D_G = |G_2 - G_1|$                   (5)

$D_B = |B_2 - B_1|$

In Eq. (5) The embedding capacity for each channel is determined based on the ranges of $D_R$, $D_G$, and $D_B$. The embedding and extraction processes remain the same as described in Section *3.3*, with the only difference being that the pixel values now correspond to the intensity values of the individual color channels. Care must be taken to ensure that the modifications made to the pixel values do not result in noticeable artifacts or color distortions.

### 3.5  Gray Image Handling

The process for handling grayscale images is similar to that for color images, with the main difference being that there is only one intensity value for each pixel instead of separate values for the RGB channels. Let $P_1$ and $P_2$ be the pixel pairs in the grayscale image [6]. The discrepancy in pixel value, denoted as D, is computed as follows: $D = |P_2 - P_1|$, The embedding capacity for the pixel pair is determined based on the range of

D. The embedding and extraction processes for grayscale images follow the same steps as described in Section 3.3. Since there is only one channel in grayscale images, the potential issues related to color distortions are not applicable in this case.

### 3.6 Benefits of Combining APVD with Pseudorandom Pixel Selection

The adaptive nature of the APVD technique, incorporating pseudorandom pixel selection can significantly enhance its performance. Pseudorandom pixel selection addresses the 'fall off boundary problem' and 'unused blocks' problem commonly encountered in PVD and APVD steganography techniques [8, 16, 18, 20]. By distributing the changes more evenly and unpredictably throughout the image, the likelihood of noticeable artifacts is reduced, and the overall visual quality of the stego-image is improved. Moreover, the pseudorandom selection of pixel pairs increases the chances of utilizing a higher number of pixel pairs for embedding, thereby enhancing the overall embedding capacity of the method. Combining APVD with pseudorandom pixel selection not only mitigates these issues but also greatly bolsters security by rendering the embedding pattern less predictable, more resistant to steganalysis, and harder for an attacker to reverse-engineer.

### 3.7 Evaluations

The effectiveness of the suggested approach was assessed by embedding secret data into the test images and extracting the data from the resulting stego-images. The effectiveness of APVD Pseudorandom Pixel Selection approach in maintaining high image quality was analyzed using the following metrics:

- PSNR (Peak Signal-to-Noise Ratio): Measures image quality; higher values are better [10, 14].

$$PSNR = 10 \times log10 \left(\frac{MAX^2}{MSE}\right) \quad (6)$$

In Eq. (6), MAX signifies the maximum pixel value, whereas MSE denotes the mean value of error squares when comparing the initial and the reformed images.

- SSIM (Structural Similarity Index): This index evaluates the structural congruence between images, with larger values reflecting lesser degradation [10, 14].

$$SSIM(x,y) = \frac{(2\mu x \mu y + C_1)(2\sigma xy + C_2)}{(\mu x^2 + \mu y^2 + C_1)(\sigma x^2 + \sigma y^2 + C_2)} \quad (7)$$

Eq. (7), where $\mu$ and $\sigma$ represent mean and standard deviation, respectively, and $C_1$, $C_2$ are constants

- UIQ (Universal Image Quality Index): Assesses image quality aspects; higher values are better. Proposed method consistently obtains higher UIQ values, demonstrating effectiveness [14], on eq. (8).

$$UIQ: UIQ(x,y) = \frac{(4\sigma xy \mu x \mu y)}{(\sigma x^2 + \sigma y^2)(\mu x^2 + \mu y^2)} \quad (8)$$

In Eq. (8), $\mu$ and $\sigma$ represent mean and standard deviation, respectively.

## 4   Data

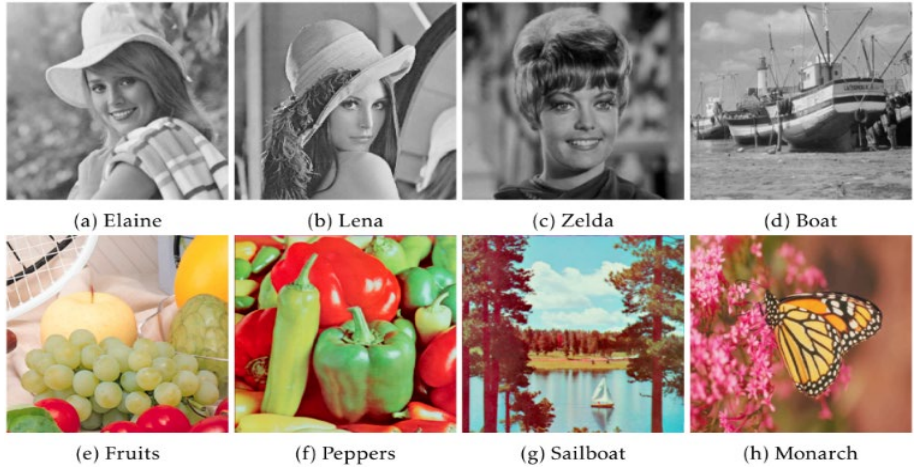

**Fig. 2.** Images of the Proposed Steganographic Process

In Fig. 2, displays a collection of test images used to evaluate the proposed steganography technique. Sizes are 256*256 Pixel. The set contains grayscale images (a to d) and color images (e to h) to showcase the efficiency of the technique for both kinds of images. The comparison between cover and stego-images highlights minimal distortion, while the secret images showcase successful data extraction. This demonstrates the versatility of the proposed technique, as it can be applied to various image formats, ensuring secure and robust communication.

## 5   Experimental Results and Evaluations Methodology

The research includes experiments with grayscale and color test images of dimensions 256x256 pixels, and alterations were made to the cover images to resize them to 512x512 pixels. Secret images were also resized to 128x128 pixels. For the grayscale tests, 'a' served as the cover image, with 'b', 'c', and 'd' as secret images. Similarly, for the color images, 'e' was the cover image, and 'f', 'g', and 'h' were the secret images. PSNR, SSIM, and UIQ values for the cover and stego images were computed for both grayscale and color sets. A results table shows these values, offering a comparison of quality and similarity between the cover, secret, and extracted images. This provides an extensive evaluation of the proposed steganography technique.

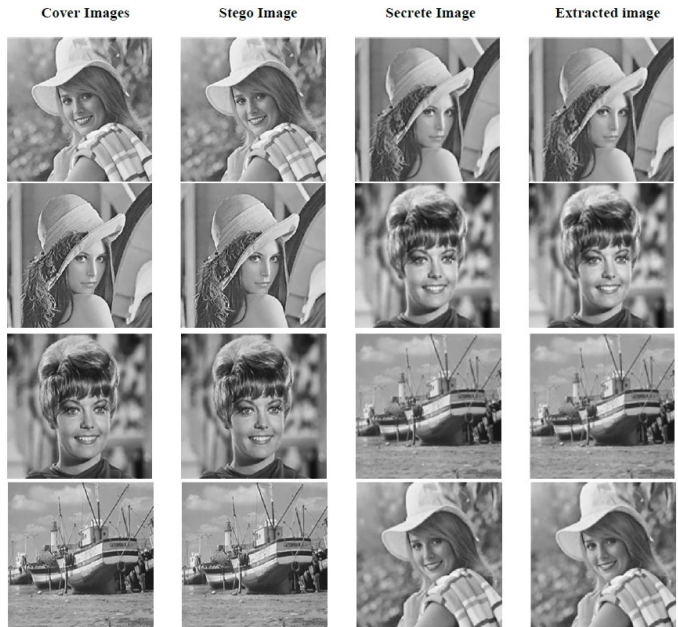

**Fig. 3.** The figure showcases four examples of visual results for grayscale images: the first column presents the cover images, the second column exhibits the stego images, the third column visually represents the secret images, and the fourth column unveils the images that contain the extracted secrets.

**Table 2.** For the gray version, assess the PSNR, SSIM, and UIQ values

| Cover Image | Secret Image | PSNR | SSIM | UIQ |
|---|---|---|---|---|
| Elaine | Leena | 54.17 | 0.9969 | 0.9981 |
| | Zelda | 54.17 | 0.9976 | 0.9973 |
| | Boat | 54.18 | 0.9958 | 0.9921 |
| Leena | Elaine | 54.14 | 0.9962 | 0.9980 |
| | Zelda | 54.14 | 0.9979 | 0.9910 |
| | Boat | 54.14 | 0.9985 | 0.9914 |
| Zelda | Elaine | 54.16 | 0.9980 | 0.9983 |
| | Leena | 54.15 | 0.9972 | 0.9922 |
| | Boat | 54.14 | 0.9951 | 0.9939 |
| Boat | Elaine | 54.13 | 0.9959 | 0.9918 |
| | Leena | 54.17 | 0.9974 | 0.9940 |
| | Zelda | 54.16 | 0.9948 | 0.9975 |

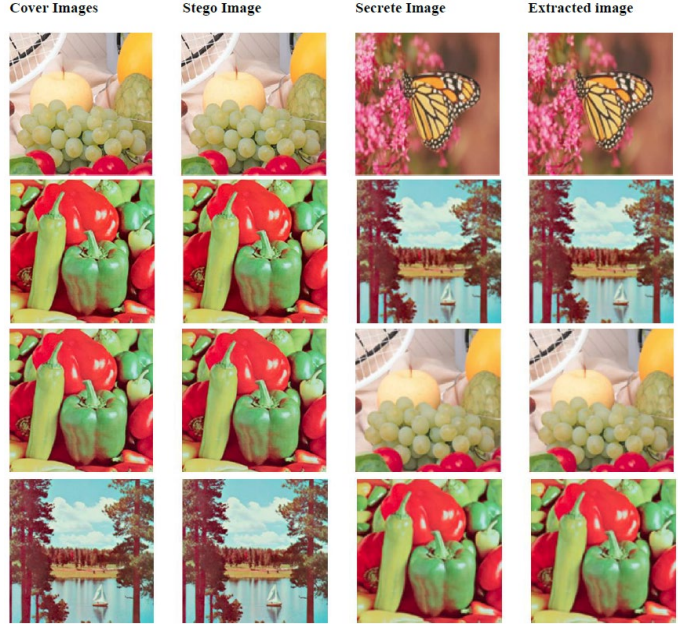

**Fig. 4.** The figure provides four instances of visual outcomes for color images: the first column introduces the cover images, the second column brings forward the stego images, the third column reveals the secret images, and the fourth column brings to light the images containing the extracted secrets.

**Table 3.** For the color version, assess the PSNR, SSIM, and UIQ values

| Cover Image | Secret Image | PSNR | SSIM | UIQ |
|---|---|---|---|---|
| Fruits | Monarch | 53.11 | 0.9967 | 0.9952 |
| | Peppers | 53.10 | 0.9920 | 0.9874 |
| | Sailboat | 53.12 | 0.9946 | 0.9810 |
| Peppers | Fruits | 53.08 | 0.9820 | 0.9950 |
| | Monarch | 53.06 | 0.9906 | 0.9917 |
| | Sailboat | 53.05 | 0.9964 | 0.9836 |
| Sailboat | Fruits | 53.03 | 0.9878 | 0.9929 |
| | Monarch | 53.01 | 0.9890 | 0.9948 |
| | Peppers | 53.02 | 0.9865 | 0.9971 |
| Monarch | Fruits | 53.07 | 0.9954 | 0.9896 |
| | Peppers | 53.04 | 0.9971 | 0.9857 |
| | Sailboat | 53.09 | 0.9869 | 0.9965 |

Experimental results, as visually depicted in Fig. 3 and Fig. 4, and quantitatively evaluated in Table 2 and Table 3, manifest the superior performance of the proposed APVD-Pseudorandom pixel selection-based steganography method over the quantum S-boxes dependent steganography method. The proposed method's excellence is highlighted by marked improvements in essential image quality metrics, namely, PSNR (a measure of the peak signal-to-noise ratio), SSIM (a gauge of structural similarity within the image), and UIQ (a universal standard for assessing image quality). Fig. 3, visually illustrates this superiority using grayscale image samples, arranged column by column showcasing cover images, stego images, secret images, and finally, the extracted secret images for easy comparative analysis. In concordance with these visual representations, Table 2 provides a quantitative evaluation, presenting the corresponding values for PSNR, SSIM, and UIQ in the context of grayscale images. Similar visual results for color images are showcased in Fig. 4, again with each column representing a different image type. The PSNR, SSIM, and UIQ values for these color images, as presented in Table 3, further corroborate proposed method's superior performance. Compared to the method described in [1] that uses quantum walks to construct S-boxes and design a steganography technique, the proposed method consistently exhibits higher PSNR, SSIM, and UIQ values for both color and grayscale images. This highlights the effectiveness of proposed method not only in maintaining superior visual quality but also in offering a higher embedding capacity and enhanced security for the hidden data.

## 6  Conclusion

This research has demonstrated that the combination of Adaptive Pixel Value Differencing (APVD) with pseudorandom pixel selection offers a highly effective image steganography technique. This novel approach provides enhanced security, embedding capacity, and visual quality preservation in contrast to established techniques, making it a promising solution for secure and reliable communication. Pseudorandom pixel selection, integrated within this technique, overcomes prevalent challenges in APVD methods, such as 'unused blocks' problem, leading to an optimized steganographic performance. Experimental outcomes highlight that the presented strategy consistently achieves elevated PSNR, SSIM, and UIQ values for both color and grayscale images, underlining its capacity to maintain visual quality while delivering robust security for the hidden data. The adaptability of this proposed methodology accommodates a broad spectrum of cover and secret images, ensuring its viability for secure communication in diverse applications and contexts.

## 7  Future Work

While the proposed APVD with pseudorandom pixel selection technique has proven effective, future research could investigate more advanced pseudorandom number generators for enhancing security. It might also be valuable to explore other embedding techniques, such as adaptive least significant bit (LSB) embedding or embedding in transform domains. The potential expansion of the proposed method to encompass other formats of digital media, such as sound recordings, visual content, and textual documents, could significantly widen its sphere of applicability. Lastly, as deep learning advances, it would be worthwhile to develop deep learning-based steganalysis methods to test the method's robustness against sophisticated detection techniques.